\documentclass[a4paper,11pt]{article}
\usepackage{pos}
\usepackage{slashed}
\usepackage{graphicx}

\newcommand{\pref}[1]{(\ref{#1})}

\newcommand{\Ep}{E_{\pi,\vec{p}}}

\newcommand{\fhat}{\hat{f}}
\newcommand{\fhateff}{\hat{f}_{\rm eff}}
\newcommand{\Dfhateff}{\Delta\fhat_{B\pi}}

\title{$B\pi$-state contamination in $B$-meson observables}
\ShortTitle{$B\pi$-state contamination in $B$-meson observables} 

\author*[a]{Oliver B{\"a}r}
\author[a]{Alexander Broll}
\author[a,b]{Rainer Sommer}

\affiliation[a]{Humboldt Universität zu Berlin,\\
  Newtonstrasse 15, 12489 Berlin, Germany}
  
\affiliation[b]{Deutsches Elektronen-Synchrotron DESY, Platanenallee 6, 15738 Zeuthen, Germany}

\emailAdd{obaer@physik.hu-berlin.de}
\emailAdd{alexander.broll@physik.hu-berlin.de}
\emailAdd{rainer.sommer@desy.de}

\abstract{Multi-particle states with additional pions are expected to result in a non-negligible excited-state contamination in lattice simulations. We show that heavy meson chiral perturbation theory can be employed to calculate the contamination due to two-particle $B\pi$ states in various $B$-meson observables like the $B$-meson decay constant and the $BB^*\pi$ coupling. We work in the static limit and to next-to-leading order in the chiral expansion. The $B\pi$ states are found to typically overestimate the observables at the few percent level depending on the size of two currently unknown NLO low-energy coefficients. A strategy to independently measure one of them with the 3-point function of the light axial vector current will be discussed.
}

\FullConference{%
 The 39th International Symposium on Lattice Field Theory,\\
  8th-13th August, 2022,\\
  Rheinische Friedrich-Wilhelms-Universit{\"a}t Bonn, Bonn, Germany
}

\begin{document}
\maketitle

\section{Introduction}\label{Intro}

Physical point simulations with the quark masses as light as in Nature eliminate the uncertainties that are associated with a chiral extrapolation of lattice data.
However, with decreasing quark masses the excited-state contamination due to multi-particle states involving light pions gets larger since the gap to the single-particle ground state shrinks with smaller pion mass.

In order to discuss this let us consider the $B$-meson 2-point (pt) function,
\begin{equation}\label{Def:C2}
C_2(t) = \sum_{\vec{x}} \langle B(\vec{x},t) B^{\dagger}(\vec{0},0)\rangle\,.
\end{equation}
Here $B$ denotes an interpolating $B$-meson field placed at source time 0 and sink time $t$. Performing the standard spectral decomposition we find (for $t > 0$ and for a finite spatial volume),
\begin{equation}\label{SpectralDecomp}
C_2(t) = b_0 e^{-M_B t} + b_1 e^{-E_1 t} + b_2 e^{-E_2 t}\, +\, \ldots \,,
\end{equation}
i.e.\ the 2pt function is a sum of exponentials with non-negative coefficients $b_k$. The leading exponential is given by the $B$-meson mass, by construction. If the heavy-light axial vector current is used as the interpolating field the leading coefficient $b_0$ is proportional to the squared $B$-meson decay constant. The remaining contribution is due to excited states with the same quantum numbers as the $B$-meson. Their exponential suppression relative to the ground state contribution is governed by the energy gap $\Delta E_k =E_k-M_B$ with $E_k$ the energy of the k-th excited state. What states actually form the first few states in the decomposition \pref{SpectralDecomp} depends strongly on the pion mass and the extent $L$ of the finite spatial box realized in the lattice simulation. For pion masses close to the physical value and for sufficiently large $L$ one expects two-particle $B^*\pi$ states with small back-to-back momentum $|\vec{p}| = 2\pi/L$ to be among the states with smallest energy $E_k$, which is, ignoring the interaction energy, approximately given by the sum of the individual $B^{\ast}$-meson and pion energies.

The impact of the excited-state contribution in \pref{SpectralDecomp} not only depends on the energy gaps but also on the coefficient ratios $b_k/b_0$. The coefficients $b_k$ are squared matrix elements involving the excited states. In case of multi-particle states with additional pions these coefficients can be computed in chiral perturbation theory (ChPT), as was first pointed out in Refs.\ \cite{Tiburzi:2009zp,Bar:2012ce}. 

By now there exist quite a few ChPT results for the $N\pi$ excited-state contamination in various nucleon correlation functions and physical observables derived from them: the nucleon mass \cite{Bar:2015zwa}, moments of parton distribution functions \cite{Bar:2016jof}, nucleon charges \cite{Bar:2016uoj} as well as nucleon form factors \cite{Bar:2018xyi,Bar:2019gfx,Bar:2019igf,Bar:2021crj}. The common framework for these calculations is covariant baryon ChPT (BChPT), and the calculations were carried out to leading order (LO). For a  review of the setup see  \cite{Bar:2017kxh}. 

Here and in  \cite{Broll:Lattice22} we report about our ongoing project to carry out analogous ChPT calculations for the phenomenologically relevant $B$-meson sector. A full account will appear elsewhere \cite{BBS1}. The motivation and main idea for this study are the same as for the nucleon sector, only the details of the low-energy effective theory change: Instead of BChPT the theoretical framework for our calculations is heavy meson ChPT. 

\section{Heavy meson ChPT basics}

We consider QCD with two light and mass degenerate up and down type quarks and a heavy $b$ quark. For the latter we assume the static approximation, i.e.\ we work at LO in the heavy quark effective theory (HQET). At this order the theory is constrained by the heavy quark spin symmetry (HQSS) \cite{Isgur:1989vq}.

The corresponding chiral effective theory is called heavy meson (HM) ChPT  \cite{Wise:1992hn,Burdman:1992gh}. With our assumed quark content the chiral effective theory contains three mass degenerate pions and the static $B$-mesons $B_q$, $B^*_q$. Since we assume isospin symmetry we will drop the light flavor index $q$ throughout. Because of HQSS the pseudoscalar $B$ and the vector meson $B^*$ are mass degenerate and their fields are collected in a heavy meson multiplet
\begin{equation}\label{DefH}
H=P_{\varv}\left( i B^*_{\mu} \gamma_{\mu} + i B \gamma^5\right)\,\qquad P_{\varv}= (1- i \slashed{\varv})/2\,.
\end{equation}
$P_{\varv}$ projects out the anti $B$-meson.
Even though we have given the general formula for arbitrary $\varv_{\mu}$ we only consider the case with the $B$-mesons being at rest, $\vec{\varv}=0$.

\begin{figure}[t]
\begin{center}
\includegraphics[scale=0.14]{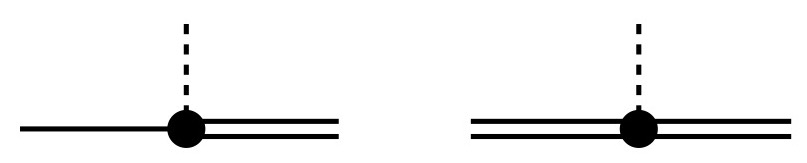}\hspace{2cm}\includegraphics[scale=0.14]{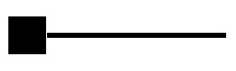}\hspace{0.5cm}
\includegraphics[scale=0.14]{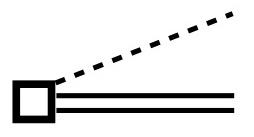}
\\[0.4ex]
\hspace{0.5cm}a) \hspace{5cm} b)\\
\caption{
a) Two LO interaction vertices coupling a pion (dashed line) with a heavy pseudoscalar (solid line) and vector meson (solid double line) and with two heavy vector mesons. Both vertices are proportional to the LO LEC $\varg$.  b) Vertices for the $B$-meson interpolating field, at LO (solid square) and NLO (open square). 
}
\label{fig:int_vertices}
\end{center}
\end{figure}

The interaction terms at LO that are relevant in the following stem from the Lagrangian
\begin{equation}
{\cal L}_{\rm int, \,LO} \,=\, i \frac{\varg}{f} \, {\rm tr}\left( H \gamma_{5} \gamma_{\mu} \partial_{\mu}\pi \bar{H} \right)\,+\,\ldots
\end{equation}
Taking the trace over the Dirac matrices it leads to the interaction vertices in figure \ref{fig:int_vertices}a. As a LO term it involves one derivative of the pion fields, i.e.\ it is proportional to the pion's four-momentum. It also involves two low-energy coefficients (LECs) $f$ and $\varg$, the chiral limit values of the pion decay constant and the $BB^*\pi$ coupling. 
%=====================
%=====================

We are interested in correlation functions of interpolating fields for the $B$-mesons. Possible interpolators at the quark level are the quark bilinears 
\begin{equation}\label{Def:Quark_bilinear}
 \overline{q}_l(x)\Gamma Q_h(x)\,,\qquad \Gamma = \gamma_5\, {\rm or}\, \gamma_k\,,
\end{equation}
with a light anti-quark $\overline{q}_l$ and a static $b$-quark $Q_h$. These are mapped to HMChPT as usual: One writes down the most general terms that transform as \pref{Def:Quark_bilinear} under the relevant symmetries. This leads to one term at LO and two independent terms at next-to-leading order (NLO). The associated vertices in Feynman graphs are displayed in figure \ref{fig:int_vertices}b in case of the interpolating field for the pseudoscalar $B$-meson. 

The chiral expressions for the interpolating fields not only hold for the local bilinears in \pref{Def:Quark_bilinear} but also for {\em smeared} interpolators with a smeared light quark field provided the smearing radius is much smaller than the Compton wave length of the pion \cite{Bar:2013ora}. The only difference between local and smeared interpolating fields are different values for the associated LECs.

In the following we present the $B\pi$ excited-state contamination in $B$-meson correlation functions through NLO in the chiral expansion.\footnote{Because of HQSS there is essentially no difference between a $B$ and a $B^{\ast}$. Therefore, we generically speak of $B\pi$ states even if the two-particle excited state involves a $B^{\ast}$.}  For this we need the interpolating fields through NLO, but only the interaction vertices from the LO chiral Lagrangian. The reason is that at NLO there are no 3-meson interaction terms like the ones given in fig.\ \ref{fig:int_vertices}a.  

\section{$B\pi$ excited-state contamination in HMChPT}

\begin{figure}[t]
\begin{center}
\includegraphics[scale=0.35]{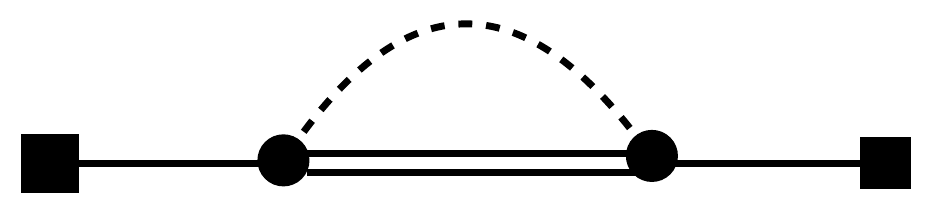}\hspace{1cm} \includegraphics[scale=0.35]{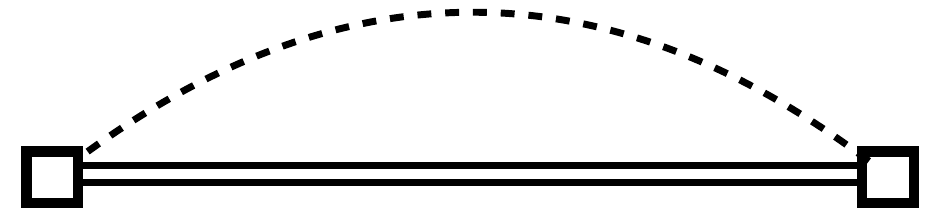}\\[1.8ex]
%a)\hspace{4cm} b)\\[2ex]
\includegraphics[scale=0.35]{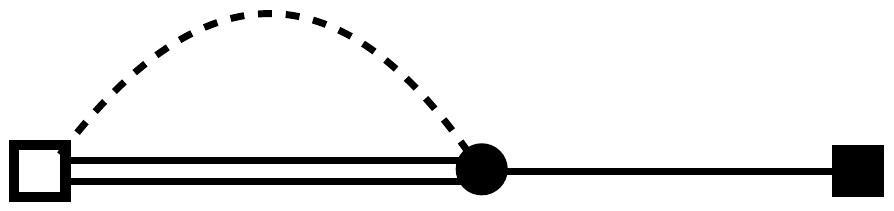}\hspace{1.1cm} \includegraphics[scale=0.35]{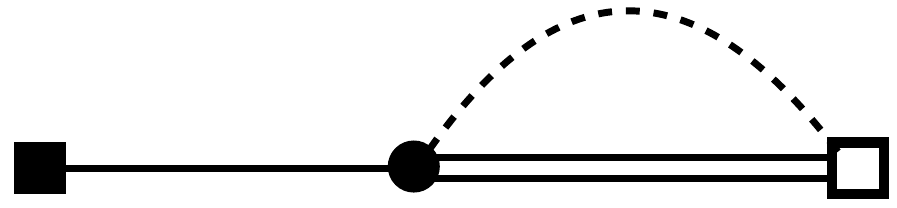}\\[0.4ex]
%c)\hspace{4cm} d)\\[2ex]
\caption{
Feynman diagrams contributing to the $B\pi$ contribution in the pseudoscalar 2pt function through NLO. The dashed and the solid double line represent pion and vector $B$-meson propagators, respectively. The remaining elements are as in fig.\ \ref{fig:int_vertices}. 
}
\label{fig:diags_C2_Bpi}
\end{center}
\end{figure}

The calculation of correlation functions is a standard task in HMChPT and amounts to the computation of Feynman diagrams. For our application we need the single $B$-meson and the $B\pi$ contribution in a given correlation function. For instance, the Feynman diagrams for the leading $B\pi$ contribution in the $B$-meson 2pt function in \pref{SpectralDecomp} are shown in figure \ref{fig:diags_C2_Bpi}. Together with the single $B$-meson contribution (one tree-level diagram, not shown) we can form the relative $B\pi$ contribution
\begin{equation}\label{Def_reldeviationC2}
\Delta C_2^{B\pi}(t) \equiv \frac{C_2^{B\pi}(t)}{C_2^{B}(t)} = \sum_{\vec{p}_k} c_{\rm 2pt}(\vec{p}_k) e^{-E_{\pi,\vec{p}_k} |t|}\,.
\end{equation}
The dimensionless coefficients $c_{\rm 2pt}(\vec{p})$ are nothing but the ratios $b_k/b_0$ introduced in section \ref{Intro} for the  case of $B\pi$ excited states, with the discrete pion momentum $\vec{p}_k$ as the excited-state label. The $c_{\rm 2pt}(\vec{p})$ are the non-trivial result of the ChPT calculation. They do not only parametrize the $B\pi$-state contamination in the 2pt function itself but also in "derived" quantities like the effective $B$-meson mass and the $B$-meson decay constant. 
%=====================
%=====================
Through NLO the explicit result for the coefficient reads
\begin{equation}\label{Resc2pt}
c_{\rm 2pt}(\vec{p}) = \frac{3}{8 (fL)^2 \Ep L } \frac{p^2}{\Ep^2} \, \left(\varg + \tilde{\beta}_1 E_{\pi,\vec{p}}   \right)^2\,.
\end{equation}
The LO result is obtained by setting the NLO LEC $\tilde{\beta}_1$ to zero. $\tilde{\beta}_1$ is one of the two NLO LECs associated with the interpolating fields for the $B$-meson. The second one, $\tilde{\beta}_2$, does not enter the pseudoscalar 2pt-function. Note that eq.\ \pref{Resc2pt} is the result for smeared interpolators at both source and sink. 
If a local interpolator is used at either source or sink it comes with a different LEC $\beta_1$, and in \pref{Resc2pt} we need to make the replacement $ \left(\varg + \tilde{\beta}_1 E_{\pi,\vec{p}}   \right)^2 \rightarrow  \left(\varg + {\beta}_1 E_{\pi,\vec{p}}   \right) \left(\varg + \tilde{\beta}_1 E_{\pi,\vec{p}}   \right)$.

\section{Impact on lattice calculations}

To LO the coefficient $c_{\rm 2pt}$ depends on three dimensionless, independent parameter combinations: $f/M_{\pi}, \varg$ and $M_{\pi}L$. The infinite volume limit of eq.\ \pref{Def_reldeviationC2} is easily taken, and in this case the number of input parameters reduces to only two. To a good approximation we can replace the LO LECs by their phenomenologically known values, $f\approx f_{\pi}\approx 93 \,{\rm MeV}$ and $\varg\approx \varg_{\pi}\approx 0.5$, and we obtain an estimate for the expected $B\pi$-state contamination as a function of the source-sink separation $t$.

The NLO LECs $\beta_1,\tilde{\beta}_1$ are not known. In section \ref{sect:ExtrLECs} we will suggest a procedure how to get estimates for them from a 3pt function involving the light axial vector current. For now we resort to a naive dimensional analysis. Both $\beta_1,\tilde{\beta}_1$ have mass dimension $-1$. In ChPT one expects this scale to be given by the chiral symmetry breaking scale $\Lambda_{\chi}^{-1}\sim \,1\, {\rm GeV}^{-1}$, and in this case we recover in \pref{Resc2pt} the standard chiral expansion in powers of $\Ep/\Lambda_{\chi}$ with dimensionless coefficients to be expected of O(1). Thus, we make the rather crude assumption
\begin{equation}\label{betarange}
-{\Lambda_{\chi}^{-1}} \le \beta_1,\tilde{\beta}_1\le {\Lambda_{\chi}^{-1}},
\end{equation}
and set $\Lambda_{\chi} = 4\pi f_{\pi} = {\rm O(1\, GeV)}$.

Having specified all input parameters the $B\pi$ contamination in the 2pt function depends on the source-sink separation $t$ only. Since ChPT is a low-energy effective theory the source-sink separation $t$ needs to be sufficiently large such that the deviation $\Delta C_2^{B\pi}(t)$ is dominated by low-energy pions with an energy $E < \Lambda_{\chi}$. It turns out that more than about 80 percent of the total $B\pi$ contribution  in eq.\ \pref{Def_reldeviationC2} stems from states with a pion energy less than 700 MeV provided the $t\gtrsim 1.3\,{\rm fm}$. So we may expect the chiral expansion to be reasonably well behaved for such source-sink separations. This is in agreement with the naive expection that euclidean time-separations need to be at least 1 fm for pion physics to dominate the correlation functions.
%======================================================================================
\begin{figure}[t]
\begin{center}
{\small $\Delta M^{B\pi}_B(t)/ [{\rm MeV}]$}\hspace{5.5cm} {\small $\Dfhateff(t)$}\\
\includegraphics[scale=0.57]{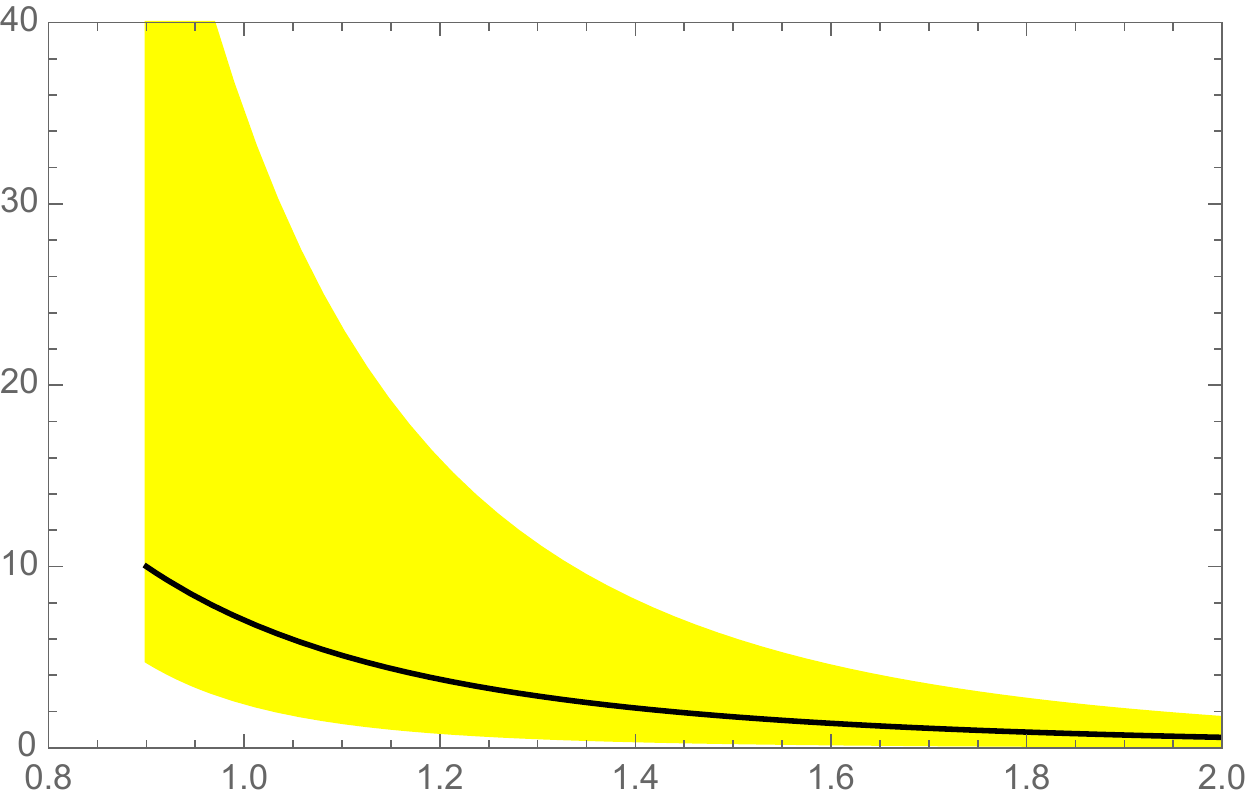}\hspace{0.5cm}\includegraphics[scale=0.57]{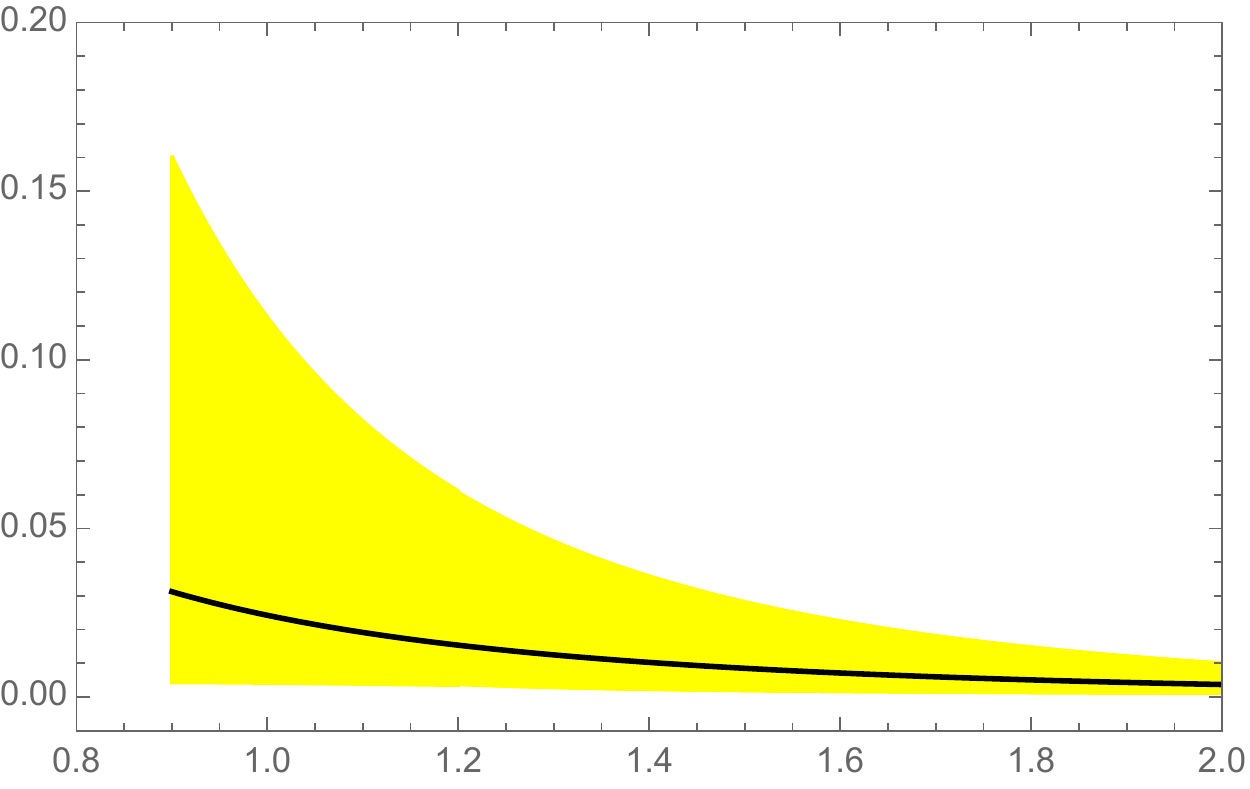}\\
{\small $t$/[fm]}\hspace{7cm}{\small$t$/[fm]}\\
\caption{The deviations $\Delta M^{B\pi}_B(t)$ in MeV (left panel) and $\Dfhateff(t)$ (right panel) as a function of the source-sink separation $t$. The solid line is the LO result, the yellow band shows the NLO result with the NLO LECs $\tilde{\beta}_1,\beta_1$  being varied as in \pref{betarange}.}
\label{fig:NImp_EffMass}
\end{center}
\end{figure}
%======================================================================================

The effective $B$-meson mass is defined as usual, $M_B^{\rm eff}(t) = -\partial_t \ln C_2(t)$. The left panel in figure \ref{fig:NImp_EffMass} shows the deviation from a constant, caused by the $B\pi$ excited states, as a function of $t$. The solid line is the LO result, and the yellow band represents the NLO result with the unknown LEC $\tilde{\beta}_1$ varied according to \pref{betarange}. At $t=1.3$ fm the NLO result varies roughly between about $+1$ and $+11$ MeV.

Analogously we can define an effective $B$-meson decay constant,
\begin{eqnarray}\label{DefFhatestimatorLS}
\fhateff(t) &=& -\frac{C^{LS}_2(t)}{\sqrt{C^{SS}_2(t)}} e^{\frac{1}{2}M_B^{\rm eff}(t)  \,t}.
\end{eqnarray}
Here $C^{LS}_2(t)$ refers to the 2-pt function with the local $A_4$ at the sink and a smeared interpolating field $B^{\dagger}$ at the source. Analogously, $C^{SS}_2(t)$ denotes the 2-pt function with smeared interpolators at both source an sink.
For finite $t$ the effective decay constant differs from the true decay constant $\fhat = f_B \sqrt{M_B}$ by a $B\pi$ contamination,
\begin{equation}
\fhateff(t) = \fhat \left(1 + \Dfhateff(t)\right)\,.
\end{equation}
The right panel in figure \ref{fig:NImp_EffMass} displays the deviation $\Dfhateff(t)$. 
As before the solid line shows the LO result, and the yellow band the NLO result with  the unknown LECs $\tilde{\beta}_1,\beta_1$ varied as discussed above. This results in $\Dfhateff(t)$ varying between $+0.5\%$ and $+5\%$ at $t=1.3$ fm. 

\section{A handle on the NLO LECs}\label{sect:ExtrLECs}

%=====================
\begin{figure}[t]
\begin{center}
\includegraphics[scale=0.14]{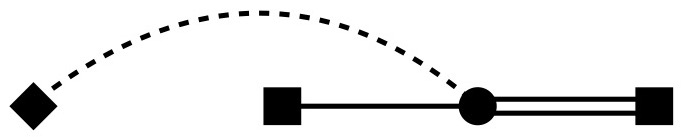}\hspace{1cm} \includegraphics[scale=0.14]{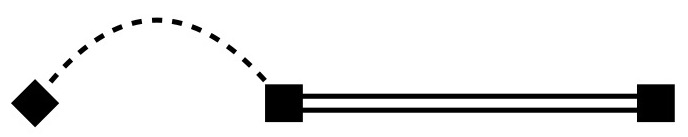}\hspace{1cm} \includegraphics[scale=0.14]{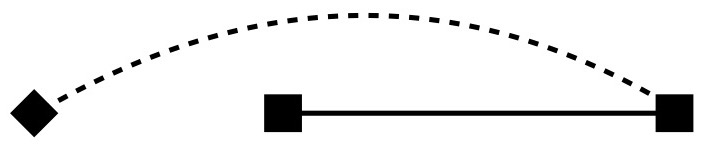}\\[0.8ex]
a)\hspace{4.2cm}b) \hspace{4.2cm} c)\\[0.4ex]
\caption{
Feynman diagrams contributing to the 3pt function in \pref{DefC3A4}. Note the time ordering $t'> t >0$ (euclidean time runs from right to left). 
}
\label{fig:diagsC3A4}
\end{center}
\end{figure}
%====================

The large yellow bands in fig.\ \ref{fig:NImp_EffMass} reflect our ignorance of the NLO LECs. Obviously, it is desirable to determine these LECs in order to have a more precise understanding about the actual  impact of the $B\pi$ states.
A straightforward and at least in principle simple procedure to measure the NLO LECs is based on the 3pt function with the timelike component $A_4$ of the light axial vector current,
\begin{equation}\label{DefC3A4}
C_{3,A_4}(\vec{q},t,t') = \int_{L^3} d^3x \int_{L^3} d^3y \, e^{i\vec{q}\cdot\vec{y}}\,\langle A_4(t',\vec{y}) B(t,\vec{x}) B^*(0,\vec{0})\rangle\,.
\end{equation}
A non-vanishing momentum transfer $\vec{q}$ is injected by the axial vector current at insertion time $t^{\prime}$. Note that it 
is placed {\em outside}  the source-sink interval by choosing $t' > t>0$. With this 3pt function we define the ratio
\begin{eqnarray}\label{DefR4ratio}
R_4(\vec{q},t,t') &\equiv &2i\frac{C_{3,A_4}(\vec{q},t,t')}{C_2(t)}\frac{C^{AA}_{4 4}(\vec{q},t)}{C^{AA}_{k 4}(\vec{q},t)}
\end{eqnarray}
with the 2pt function $C_2$ defined in \pref{Def:C2}. $C^{AA}_{\mu,4}$ denotes the 2pt function of the light axial vector currents $A_{\mu}$ and $A_4$. 
All these correlation functions are easily computed to LO in ChPT. Fig.\ \ref{fig:diagsC3A4} displays the dominant diagrams for the 3pt function in \pref{DefC3A4}, leading to
\begin{eqnarray}\label{LOresR4}
R_4(\vec{q},t,t')&=& \left(\varg-\tilde{\beta}_1 E_{\pi,q}\right) e^{-E_{\pi.q}(t'-t)} + {\rm O}\left(e^{-E_{\pi,\vec q} t'}\right)\, +\, \ldots \,,\label{genR}
\end{eqnarray}
if a smeared interpolator for the $B$-meson is employed at sink time $t$. By construction, the exponential in \pref{LOresR4} has a simple prefactor involving the LECs $\varg$ and $\tilde{\beta}_1$. The ellipses denote the excited state contribution in $R_4$ that can be ignored for our purposes.

Result \pref{LOresR4} suggests the following procedure: Determine $\exp[-E_{\pi,q}(t'-t)] R_4(\vec{q},t,t') = \varg-\tilde{\beta}_1 E_{\pi,q}$. 
If an estimate for $\varg$ is known this determines $\tilde{\beta}_1$, otherwise (at least) two values of $\vec{q}$ are needed. In principle this procedure works for both local and smeared interpolating fields, thus both $\tilde{\beta}_1$ and $\beta_1$ are accessible by the same procedure. Whether it works in practice is a separate question and needs to be checked with actual lattice simulations.

\section{Excited-state contamination in the $BB^*\pi$ coupling}

Besides the effective $B$-meson mass and decay constant the determination of the $BB^*\pi$ coupling itself is afflicted with the $B\pi$ excited state contamination.  
The $BB^*\pi$ coupling $\varg_{\pi}$  is usually obtained from the 3pt function with a spatial component of the light axial vector current.\footnote{Recall: the LEC $\varg$ that we have encountered so far is the chiral limit value of $\varg_{\pi}$.} It is essentially the 3pt function in \pref{DefC3A4} but with $\vec{q}=0$ and the current $A_k$ placed between source and sink. 
With the 3pt function the ratio with the $B$-meson 2pt function and the summation estimate are formed, see \cite{Bernardoni:2014kla}, for instance. It depends on the source-sink separation $t$ and is of the general form
\begin{equation}
\varg_{\pi}^{\rm sum}(t)= \varg_{\pi}\Big(1+\Delta \varg^{{\rm sum}}_{\pi}(t)\Big)\,,
\end{equation}
where $\Delta \varg^{{\rm sum}}_{\pi}(t)$ accounts for the excited state contribution that is present for finite $t$.

The HMChPT calculation of the $B\pi$ contribution in $\varg^{\rm sum}_{\pi}(t)$ is analogous to what we have described so far. Through NLO twelve diagrams need to be computed. At NLO one more LEC, stemming from the NLO expression of the light axial vector current, enters the ChPT result and adds another unknown parameter. It is not as easily determined as $\tilde{\beta}_1,\beta_1$, a 4pt function seems to be required as a handle to it.

%=============================================================================
\begin{figure}[tbp]
\begin{center}
{\small $\Delta \varg^{{\rm sum}}_{\pi}$}\\
\includegraphics[scale=0.55]{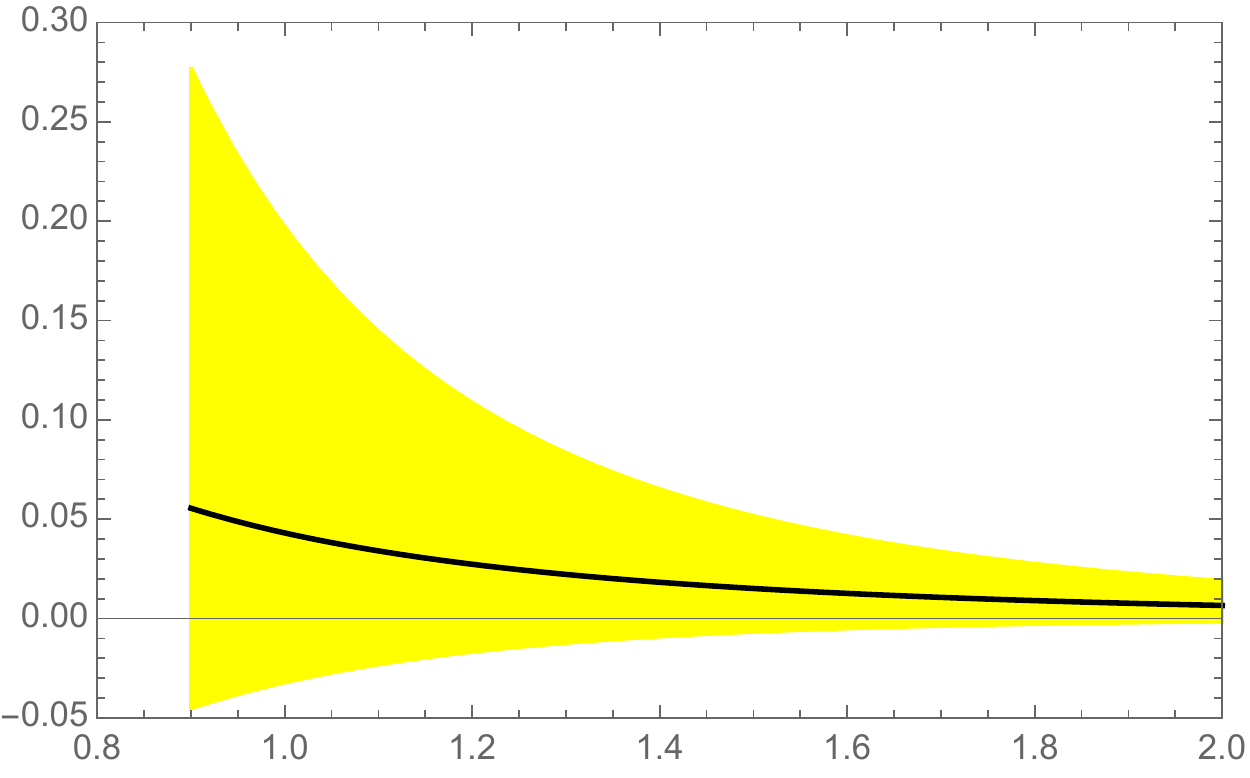}\\[2pt]
{\small $t$/[fm]}\\
\caption{The deviation $\Delta \varg^{{\rm sum}}_{\pi}$ as a function of the source-sink separation $t$. The solid line is the LO result, the yellow band shows the NLO result with the NLO LECs being varied as in \pref{betarange}.}
\label{fig:NImp_gsum}
\end{center}
\end{figure}
%=============================================================================
Figure \ref{fig:NImp_gsum} shows the deviation $\Delta \varg^{{\rm sum}}_{\pi}(t)$ as a function of $t$. As before the solid line shows the LO result, the yellow band is the NLO result with all NLO LECs being varied according to \pref{betarange}. At $t=1.3$ fm the $B\pi$ state contamination leads to a misestimation between $-1\% \ldots +8\%$. This spread is reduced by about a factor 1/2 if the NLO LEC $\tilde{\beta}_1$ is known.

\section{Conclusions}

ChPT can be used to obtain estimates for the excited-state contamination in $B$-meson observables due to two-particle $B\pi$ states. Working to NLO the $B\pi$ contamination in the $B$-meson decay constant and the $BB^*\pi$ coupling is roughly at the five-percent level for source-sink separations of about $1.3$ fm.

The predictive power of the ChPT results is limited by our lack of knowledge of the NLO LECs associated with the $B$-meson interpolating fields. 
The next natural step to do is to determine these LECs as we described in section \ref{sect:ExtrLECs}. It is particularly interesting to study how a concrete smearing procedure affects the value of the LEC and, subsequently,  the size of the $B\pi$ contamination. A smearing is of course expected to reduce the excited-state contamination in the correlation functions, but how big this effect is on the $B\pi$ contribution is not understood.
 
We have computed the $B\pi$ contribution to NLO in the chiral expansion. In principle it is possible to go to the next chiral order, however, it is doubtful whether this is useful in practice. Additional LECs will enter the results, and we expect most of them to be not easily accessible.

More interesting is the computation of the $B\pi$ contamination in other $B$-physics observables, for instance the form factors relevant for the semi-leptonic decay $B\rightarrow\pi$ and the determination of the CKM matrix element $V_{ub}$. This is the topic of A.\ Broll's contribution to this conference \cite{Broll:Lattice22}.

\vspace{4ex}
\noindent {\bf Acknowledgments}
AB’s research is funded by the Deutsche Forschungsgemeinschaft (DFG, German Research Foundation), Projektnummer 417533893/GRK2575 “Rethinking Quantum Field Theory”.
\vspace{3ex}

\providecommand{\href}[2]{#2}\begingroup\raggedright\endgroup

\end{document}